\documentclass[12pt,preprint]{aastex}
\usepackage{amsmath}

\shorttitle{Disrupted asteroid P/2016 G1}
\shortauthors{Moreno et al.}

\begin{document}


\title{Disrupted asteroid P/2016 G1. II. Follow-up observations from
  the Hubble Space Telescope}


\author{F. Moreno\affil{Instituto de Astrof\'\i sica de Andaluc\'\i a, CSIC,
  Glorieta de la Astronom\'\i a s/n, 18008 Granada, Spain}
\email{fernando@iaa.es}}

\author{
J. Licandro\affil{Instituto de Astrof\'\i sica de Canarias,
  c/V\'{\i}a 
L\'actea s/n, 38200 La Laguna, Tenerife, Spain, 
\and 
 Departamento de Astrof\'{\i}sica, Universidad de
  La Laguna (ULL), E-38205 La Laguna, Tenerife, Spain}}  

\author{
M. Mutchler\affil{
Space Telescope Science Institute, 3700 San Martin Drive, Baltimore,
MD 21218, USA}}

\author{
A. Cabrera-Lavers\affil{Instituto de Astrof\'\i sica de Canarias,
  c/V\'{\i}a 
L\'actea s/n, 38200 La Laguna, Tenerife, Spain, 
\and 
GRANTECAN, Cuesta de San Jos\'e s/n, E-38712 , Bre\~na Baja, La Palma,
Spain}}

\author{N. Pinilla-Alonso\affil{
Florida Space Institute, University of Central Florida, Orlando, FL
32816, USA}}

\and 

\author{F.J. Pozuelos\affil{Instituto de Astrof\'\i sica de Andaluc\'\i a, CSIC,
  Glorieta de la Astronom\'\i a s/n, 18008 Granada, Spain, 
\and
Space Sciences, Technologies and Astrophysics Research (STAR)
Institute, Universit\'e de Li\`ege, all\'ee du 6 Ao\^ut 17,  Li\`ege,
Belgique. }  }


\begin{abstract}

After the early observations of the disrupted asteroid P/2016 G1 with the
10.4m Gran Telescopio Canarias (GTC), and the modeling of the dust
ejecta, we have performed a follow-up observational campaign of this
object using the Hubble Space Telescope (HST) during two epochs (June 28 and
July 11, 2016). The analysis of these HST images with the same model
inputs obtained from the GTC images revealed a good consistency
with the predicted evolution from the GTC images, 
so that the model is applicable to the whole
observational period from late April to early July 2016. This result
confirms that 
the resulting dust ejecta was caused by a relatively short-duration
event with onset about 350 days before perihelion, and spanning about 30 days
(HWHM). For a size distribution of particles with a geometric albedo
of 0.15, having radii limits of 1
$\mu$m and 1 cm, and following a power-law with index --3.0, the total
dust mass ejected is $\sim$2$\times$10$^7$ kg. As was the case with
the GTC observations, no condensations in the images that could be
attributed to a nucleus or fragments released after the disruption event
were found. However, the higher limiting
magnitude reachable with the HST images in comparison with those from GTC 
allowed us to impose a more stringent upper limit to the observed
fragments of $\sim$30 m.

\end{abstract}


\section{Introduction}

Asteroid P/2016 G1 (Panstarrs) was discovered by R. Weryk and R. J. Wainscoat 
on CCD images acquired on 2016 April 1 UT with the 
1.8-m Pan-STARRS1 telescope \citep{Weryk16}. Its Tisserand parameter
with respect to Jupiter \citep{Kresak82} can be calculated as $T_J$=3.38,
so that the object belongs dynamically to the main asteroid belt, yet
showing cometary appearance.  The first object of this type, and
  the best characterized so far, was
  discovered by E.W. Elst and G. Pizarro \citep{Elst96}, currently
  designated as 133P/Elst-Pizarro. This object constitutes the target of a
  proposed European 
Space Agency Mission called Castalia \citep{Snodgrass17}. This new
class of objects in the Solar System today comprises about twenty
members. The proposed activation
mechanisms for these objects are very diverse, ranging from
sublimation-driven to rotational instabilities.  
\cite{Jewitt15} give an excellent review of the
different objects discovered so far, their orbital stability,  and
their activation mechanisms.

In a previous paper \citep[Paper I,][]{Moreno16}, 
disrupted asteroid P/2016 G1 was observed with
instrumentation attached to the GTC, from late April to early June,
2016, and a Monte Carlo dust tail model
of the ejecta was applied to obtain the dust physical properties.     
In this paper we report follow-up observations of P/2016 G1 acquired with the
HST, during two epochs (late June and early July, 2016) and apply the 
same dust model to the images, to assess the validity of the model
parameters over a longer temporal baseline.

\section{Observations and data reduction}

\begin{figure}[ht]
\centerline{\includegraphics[scale=0.5,angle=-90]{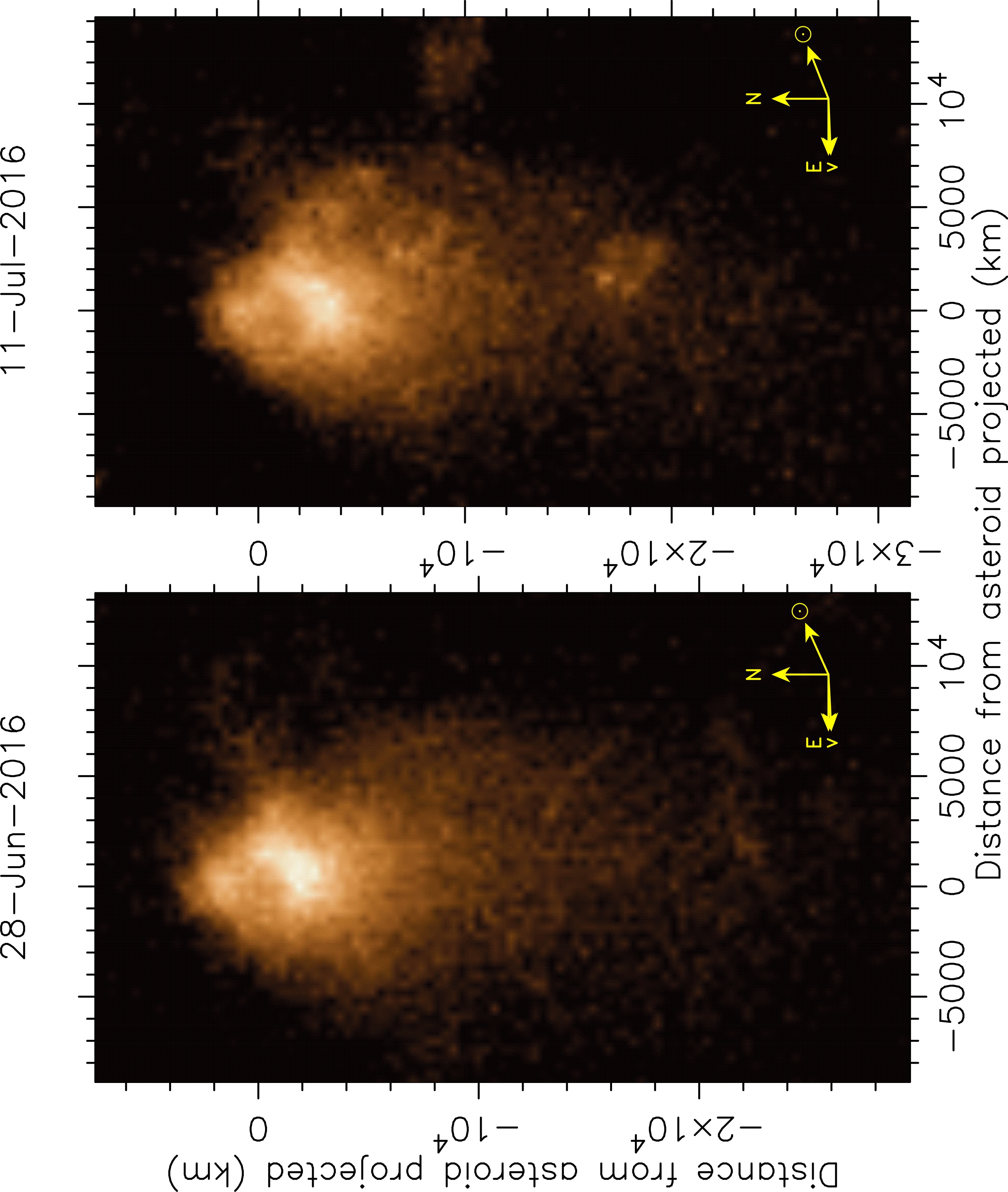}}
\caption{Drizzle-combined and 8$\times$ binned images of P/2016 G1
  obtained with WFC3 with the F350LP filter on the Hubble Space
  Telescope at the dates 
  shown.  The direction of the celestial North and East are
  indicated together with the Sun and the orbital velocity
  motion. Artifacts caused by imperfect removal of bright stars or galaxies
  are seen at coordinates (2500,--18000) and (12000,--9000) in the July
  11 image.
   \label{fig:fig1}}
\end{figure}

Observations of P/2016 G1 were performed using the Wide Field 
Camera 3 (WFC3) with the wide-band filter F350LP, which has
an effective wavelength of 584.6 nm and a width of 475.8 nm. The
object was observed on two epochs, June 28, and July 11,
2016. In the first observing run, 5 frames of 420 s exposure time
each, were acquired. In the second, 4 images were taken, of 580 s
exposure time each. Table 1 shows the log of the observations, where
the observation times refer to the starting UT time at each observing
date. Table 1 displays the relevant geometric parameters, namely the
geocentric ($\Delta$) and heliocentric ($R$) distances, the phase
angle ($\alpha$), the position angle of the extended Sun-to-asteroid radius
vector ($PsAng$), and the angle between the Earth and target orbital plane
($PlAng$), this latter parameter showing values close to the latest
GTC observation on 
June 8, 2016, so that the appearance of the object is similar to the
GTC image of that date (see Paper I, figure 1, and compare with
Fig.~\ref{fig:fig1}). 

Our target was placed on the UVIS2 chip of WFC3, which provides an
image scale of 0.04$\arcsec$/px, giving pixel sizes of 46 km and 49 km at the
asteroid on 28 June 2016 and 11 July 2016, respectively. At each
epoch,  our dithered images were median-combined, resulting in the
rejection of background sources, cosmic rays, and bad pixels. 
As the object was so faint, 
the resulting combined images were binned 8$\times$ in order to increase signal
to noise ratio which becomes $\sim$10 when averaged over the
  object. Since no nucleus or  
other spatial reference exists, the alignment procedure of the images
was difficult, so that some blurring cannot be ruled out.       

For the F350LP filter, we used a flux calibration factor (the so-called
PHOTFLAM parameter) of 5.297$\times$10$^{-20}$ erg cm$^{-2}$ s$^{-1}$
\AA$^{-1}$. The calibration formula, relating the $C=DN s^{-1}$ values of the
image to surface brightness $S$ is given by
$S=(PHOTFLAM/\Omega) C$, where $\Omega$=3.76$\times$10$^{-14}$ sr is
the solid angle subtented by a pixel. The surface brightness values are finally
converted to solar disk intensity units ($i/i_0$), which are the output units of
the Monte Carlo dust tail code. To do this, we need to convolve the
intensity of the mean solar disk spectrum \citep{Cox00} with the
filter response, which gives $i_0$=2.23$\times$10$^{6}$ erg cm$^{-2}$ s$^{-1}$
\AA$^{-1}$ sr$^{-1}$. We finally get $i/i_0 = 6.31\times 10^{-13} S$.

The resulting combined images of the asteroid are shown in
Fig.~\ref{fig:fig1}. It is apparent the lack of any condensation that
could be attributed to a nucleus or a fragment that could result from
a disruption event. The object presents a very diffuse structure,
becoming much more diluted than in the images taken previously with
the GTC (Paper I), although retaining the main features, the inverted C-shaped
structure near the head with an inner darker region near the predicted
nucleus position relative to the
surrounding material, and a slight westward lobe, more clearly
apparent in the June 28 image. The increasing diffuseness is clearly a
consequence of the expected outward expansion of the disrupted material and
the effect of radiation pressure. This will be tested in the modeling
of the ejecta as described in the next section.

As a result of the disruption event, in Paper I we found no fragments
larger than $\sim$50m (assuming a  
geometric albedo of $p_v$=0.15). In these observations the 
limiting magnitude (considering a signal-to-noise ratio S/N=3) of 
the HST combined images obtained with 
the F350LP in both observing dates would be $V\sim$27.2 for a G2V star
according to the HST exposure time calculator 
(http://etc.stsci.edu/etc/input/wfc3uvis/imaging/). We compute
  the absolute magnitudes as $H_v=V-5\log(R \Delta)-\beta\alpha$
where $\beta$=0.03 mag deg$^{-1}$ is the adopted linear phase
coefficient. For the geometric conditions of the observations (see Table 1), 
the absolute 
magnitudes become H$_v$=23.7 and H$_v$=23.5 on June 28
and July 11, respectively. Adopting the empirical
  equation $D=\frac{1329}{\sqrt{p_v}}10^{-H_v/5}$ \citep{Harris02}, 
  where $D$ is the fragment diameter, those absolute magnitudes 
  can be translated into fragment 
  radii of 31 m and 34 m, respectively. Hence, no fragments larger 
than $\sim$30 m in radius would remain from the asteroid disruption
that produced 
the observed activity. However, since the limiting magnitude refers to a dark
background and not to a source located within a faint coma, this is
actually an optimistic size limit. 

\section{The Model}

The Monte Carlo dust tail model is described in Paper I, and will not
be repeated here. The input parameters of the model were the dust-loss
rate as a function of time, and the particle velocities. A
half-Gaussian function was adopted for the dust loss rate, which is
defined by a peak activity $\dot{M}_0$,  located
at the event onset $t_0$, and having a half-width at
half-maximum denoted by HWHM, which is a measure of the effective time
span of the event. For the particle velocities, we adopted a random
function of the form $v = v_1 + \zeta v_2$, 
where $\zeta$ is a random number in the $[0,1]$ interval, and $v_1$
and $v_2$ were the fitting parameters. The remaining dust parameters
were set to the following values: the particles were assumed to be
distributed in a broad range of sizes from r=1 $\mu$m to r=1 cm, and
following a power-law differential size distribution with index
$\kappa$=--3. The particles had a density of 
$\rho_p$=1000 kg m$^{-3}$, and a geometric albedo of $p_v$=0.15. The
phase function correction was 
 performed using a linear phase coefficient of $\beta$=0.03 mag
 deg$^{-1}$, as given above, which is in the range of comet dust
 particles in the 
1$^\circ \le \alpha \le$ 30$^\circ$ phase angle domain \citep{MeechJewitt87}.  
The particle ejection was assumed isotropic, except at the beginning
of the event, where it was set to occur along a privileged direction for a
short time interval. This ad-hoc assumption was made in order to
explain the westward extension on the head, very clearly seen in
the GTC images (see Paper I, figure 4), but very diffuse on these HST
images. This direction 
was found to be given by $u_r \sim$0.98, $u_\theta \sim$0.18,
$u_z\sim$0.08, where $(u_r,u_\theta,u_z)$ are unit vectors defining a
cometocentric reference system, with $u_r$ pointing away from 
the Sun, $u_\theta$  is perpendicular to $u_r$ in the orbital plane
and opposite to the comet motion, and $u_z$ is perpendicular to the
orbital plane. 

\begin{figure}[ht]
\centerline{\includegraphics[scale=0.5,angle=-90]{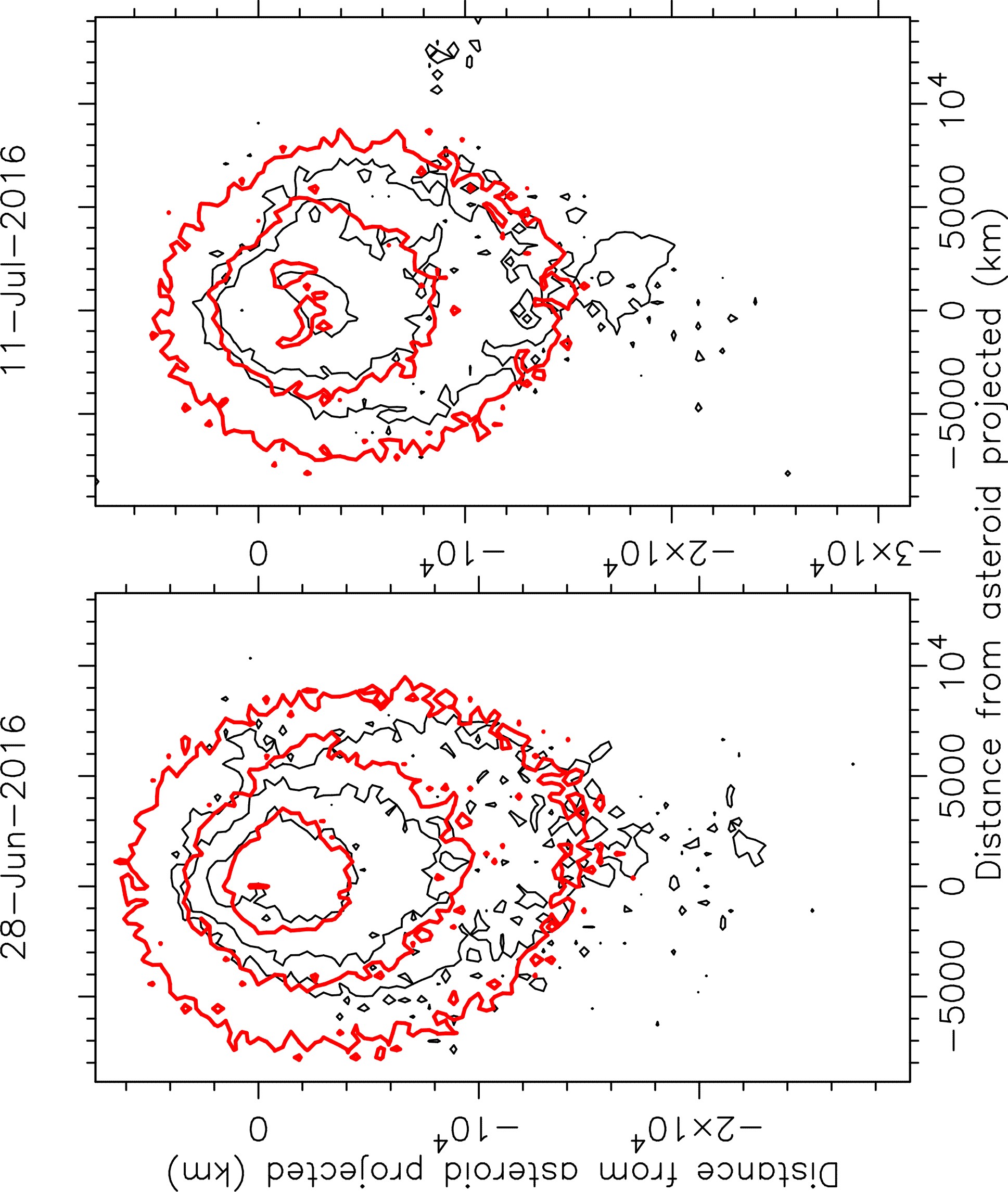}}
\caption{Measured (black contours) and modeled (thick red contours)
  isophotes for the two 
  epochs as indicated. Innermost isophote levels correspond to
  1.4$\times$10$^{-14}$ solar disk intensity units, 
and decrease in factors of two outwards. As in Fig. 1, North is up,
and East to the left. 
   \label{fig:fig2}}
\end{figure}

The application of those input model parameters to the HST images
resulted in a good agreement, although the fits are still improved by
increasing the peak dust loss rate from 7.6 kg s$^{-1}$ to 10.5 kg
s$^{-1}$. Fig.~\ref{fig:fig2} displays the measured and modeled
contours, showing an excellent agreement. This demonstrates that the
model parameters derived from the previous GTC images are compatible 
with the follow-up HST images, thereby providing stronger confidence
on the validity of the results.

\section{Conclusions}  

From the combined GTC and follow-up HST observations and the application of the 
dust tail modeling to the disrupted asteroid P/2016 G1, the following
conclusions can be drawn:

1) Asteroid P/2016 G1 was activated 350$^{+10}_{-30}$ days before
perihelion, i.e., around 10th February 2016. The activity
had a duration of 24$^{+10}_{-{7}}$ days
(HWHM). The total dust mass emitted was 
at least $\sim$2$\times$10$^7$ kg, 
with a maximum level of activity of $\sim$8-11 kg s$^{-1}$. The
dust loss mass rate  
parameters were estimated assuming a power-law size distribution of
particles between 1 $\mu$m and 1 cm, with power index of
$\kappa$=--3.0, geometric albedo of 0.15, 
and being emitted isotropically.

2) The isotropic ejection model is able to reproduce approximately the
observed tail evolution of the disrupted target. An impulsive,
short-duration ejection event in a privileged direction pointing
approximately away from the Sun has been invoked to produce a dust
feature near the head of the asteroid in the west direction. We
hypothesize that this could be atributted to an impact that
triggered the disruption of the asteroid.

3) The ejection velocities inferred from the dust model  
are very small, being in the range from 0.015 to 
0.14 m s$^{-1}$, with an average value of $\sim$0.08 m s$^{-1}$,
corresponding to the escape velocity of an object of 35 m radius and
3000 kg m$^{-3}$ density. On the other hand, from the HST
observations, we find that the upper limit to the fragment sizes in
the image is $\sim$30 m. These results are consistent, and explain
why we were not able to find any sizable fragment.

\acknowledgments

We are indebted to an anonymous reviewer for his/her useful comments
that allowed us to improve our paper.

This paper is based on observations made with the NASA/ESA Hubble
Space Telescope, 
obtained at the Space Telescope Science
Institute, which is operated by the Association of Universities for
Research in Astronomy, Inc., under NASA contract NAS 5-26555. These
observations are associated with program \#14524.”

This work was supported by contracts AYA2015-67152-R and
AYA2015-71975-REDT from the Spanish
Ministerio de Econom\'\i a y Competitividad (MINECO,
Spain). J.L. gratefully acknowledges support from 
contract AYA2015-67772-R (MINECO, Spain). N.P.-A.  
acknowledges support for Program number HST-GO-14524 provided
by NASA through a grant from the Space Telescope Science Institute,
which is operated by the Association of Universities for Research in
Astronomy, Incorporated, under NASA contract NAS5-26555. F.J.P. 
is supported by Marie Curie CO-FUND fellowship, co-founded by the University of
Li\`ege and the European Union.

\clearpage

\clearpage

\clearpage

\begin{deluxetable}{cccccccc}
\tablewidth{0pt}
\tablecaption{Log of the observations}
\tablehead{
\colhead{Start UT} & \colhead{Days to} & Total &  
\colhead{R} & \colhead{$\Delta$} &  \colhead{$\alpha$}&  \colhead{PsAng}
& \colhead{PlAng} \\
\colhead{YYYY/MM/DD HH:MM} & \colhead{perihelion} &  \colhead{exp. time (s)} &  
\colhead{(AU)} & \colhead{(AU)} &  \colhead{($^\circ$)}&  \colhead{($^\circ$)}
& \colhead{($^\circ$)} \\
}
\startdata
2016/06/28 14:53 & --211.6 &  2100 & 2.318 & 1.592 & 21.3 & 114.28 & -6.67\\
2016/07/11 10:42 & --198.8 &  2320 & 2.290 & 1.698 & 24.2 & 114.45 & -6.52\\
\enddata
\end{deluxetable}

\end{document}